\documentclass[10pt,oneside,reqno]{amsart}
\usepackage{cite}
\usepackage[%
font={small,sf},
labelfont=bf,
format=hang,    
format=plain,
margin=0pt,
width=0.8\textwidth,
]{caption}
\usepackage[list=true]{subcaption}

\usepackage{graphicx}

\begin{document}

\title{1-Multisoliton and other invariant solutions of combined KdV-nKdV  equation by using Lie Symmetry Approach}


\author{Sachin Kumar
}
\address{Department of Mathematics, University of Delhi\\
	Delhi -110007,
	India}
\email{sachinambariya@gmail.com}

\author{Dharmendra Kumar
}
\address{Department of Mathematics, SGTB Khalsa College, University of Delhi-110007, India}
\email{dharmendrakumar@sgtbkhalsa.du.ac.in}





\begin{abstract}
Lie symmetry method is applied to investigate symmetries of the combined KdV-nKdV equation, that is a new integrable equation by combining the KdV equation and negative order KdV equation. Symmetries which are obtained in this article, are further helpful for reducing the combined KdV-nKdV equation into ordinary differential equation. Moreover, a set of eight invariant solutions for combined KdV-nKdV equation is obtained by using proposed method. 
Out of the eight solutions so obtained in which two solutions generate progressive wave solutions, five are singular solutions and one multisoliton solutions which is in terms of WeierstrassZeta function.
\end{abstract}

\maketitle


\section{Introduction}
Korteweg and Vries derived KdV equation \cite{KdV1895} to model Russell's phenomenon of solitons like shallow water waves with small but finite
amplitudes \cite{wazzan}. Solitons are localized waves that
propagate without change of it's shape and velocity
properties and stable against mutual collision \cite{khattak,WazwazSWT}. It
has also been used to describe a number of important physical phenomena such as magneto hydrodynamics waves in warm plasma, acoustic
waves in an in-harmonic crystal and ion-acoustic
waves \cite{oziz2006,Brauer,khuriKdV}.
A special class of analytical solutions of KdV equation, the so-called traveling waves, for nonlinear evolution equations (NEEs) is of fundamental importance because a lot of mathematical-physical
models are often described by such a wave phenomena. Thus, the investigation of traveling wave solutions is becoming more
and more attractive in nonlinear science nowadays. However, not all equations posed in these fields are solvable. As a result,
many new techniques have been successfully developed by a diverse group of mathematicians and physicists, such as Rational function method \cite{mawa3,mawa4},
B\"acklund transformation method \cite{Backlund}, Hirota’s bilinear method \cite{Hirota,mao}, Lie symmetry method\cite{STM,zhao,ray,sachin1,sachin2,mukesh1}, Jacobi elliptic function method \cite{Jacobi}, Sine-cosine function method \cite{sinecosine}, Tanh-coth function method \cite{tanhcoth}, Weierstrass function method \cite{weierstrass}, Homogeneous balance method  \cite{homogeneousbalancemethod}, Exp-function method \cite{expfunctionmethod}, $(G'/G)$-expansion method \cite{G'/G}, etc. But, it is extremely difficult and time consuming to solve nonlinear problems with the well-known traditional methods.
This work investigates the combined KdV-nKdV equation
\begin{equation}\label{eq1}
u_{xt} +6 u_x u_{xx} +u_{xxxx}+u_{xxxt}+4u_x u_{xt}+2 u_{xx} u_t=0,
\end{equation}
where $u=u(x,t)$. We apply Lie symmetry analysis on combined KdV-nKdV equation, first constructed by Wazwaz \cite{wazwazKdVnKdV} using recursion operator \cite{Verosky}. In addition, the combined KdV-nKdV equation \eqref{eq1} possesses the Painlev\'{e} property for complete integrability \cite{Fokas1987}.
In this paper, Lie point symmetry generators of the combined KdV-nKdV equation were derived. Similarity reductions and number of explicit invariant solutions for the  equation using Lie symmetry method were obtained. All the new invariant solutions of combined KdV-nKdV were analyzed graphically. Also, 1-multisoliton solution obtained in terms of  WeierstrassZeta function which appear in classical mechanics such as, motion in cubic and quartic potentials, description of the movement of a spherical pendulum, and in construction of minimal
surfaces  \cite{wolfram}. Some of the outcomes are interesting in physical sciences and are beautiful in mathematical sciences. 

\begin{figure}
	\centering
	\subcaptionbox{Progressive wave shown by Eq. \eqref{u1awithc}}{\includegraphics[width=0.480\textwidth]{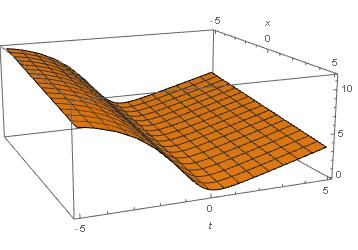}}%
	\hfill
	\subcaptionbox{Singularity on $1+x-t=0$ in Eq. \eqref{u1bwithc}}{\includegraphics[width=0.480\textwidth]{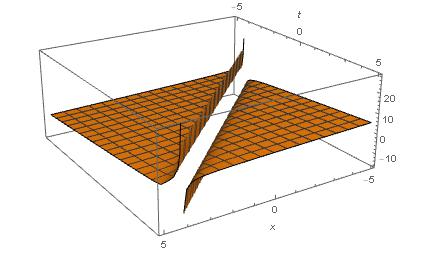}}%
	\hfill 
	
	\caption{Invariant solution profiles for Eq. \eqref{u1awithc} and Eq. \eqref{u1bwithc}}
	\label{f1a}
\end{figure}

\begin{figure}
	\centering
	\subcaptionbox{}{\includegraphics[width=0.50\textwidth]{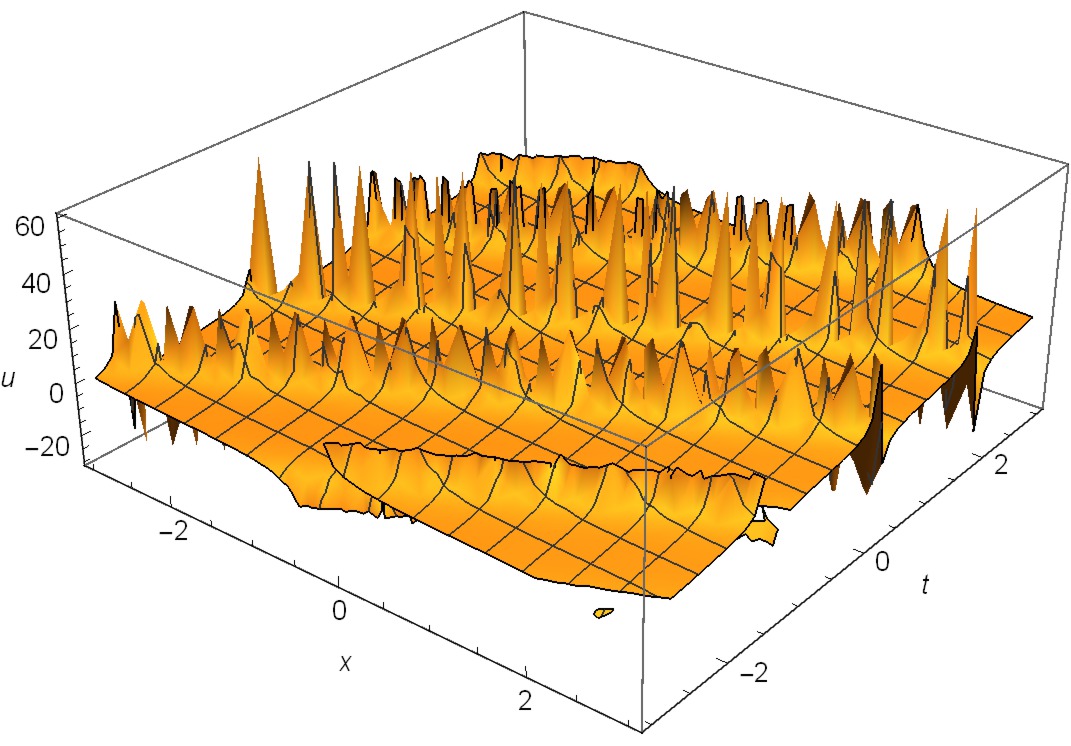}}%
	\hfill
	\subcaptionbox{}{\includegraphics[width=0.50\textwidth]{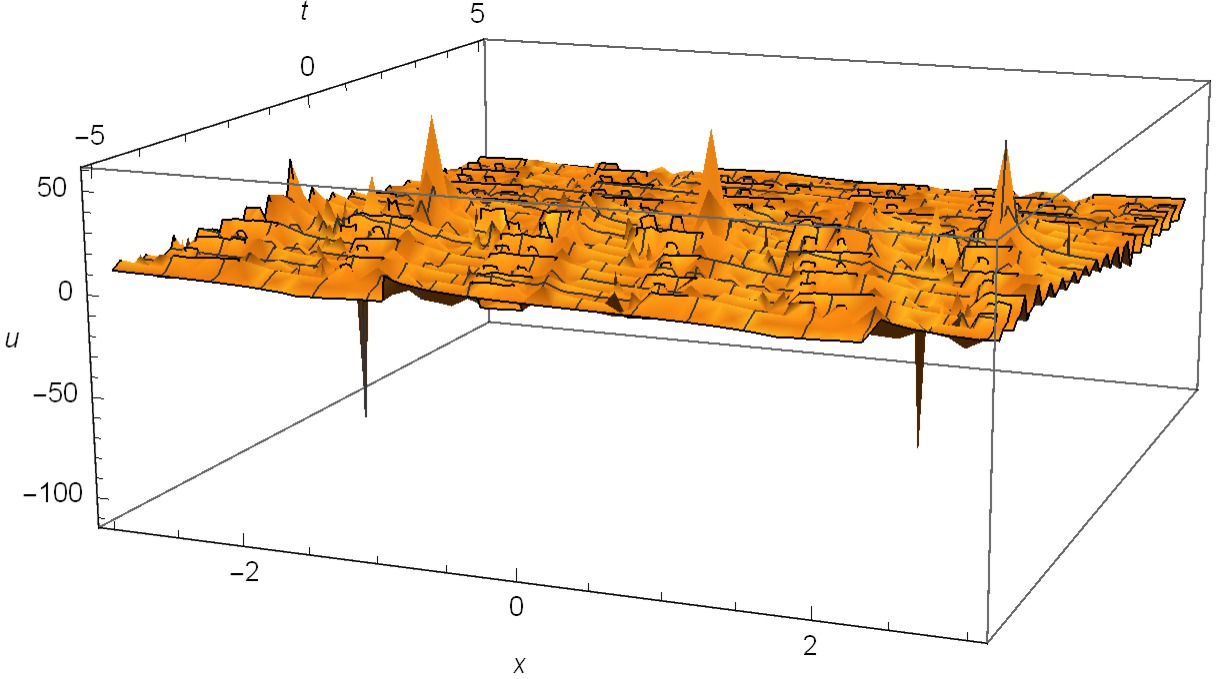}}
	\hfill \\
	\subcaptionbox{}{\includegraphics[width=0.50\textwidth]{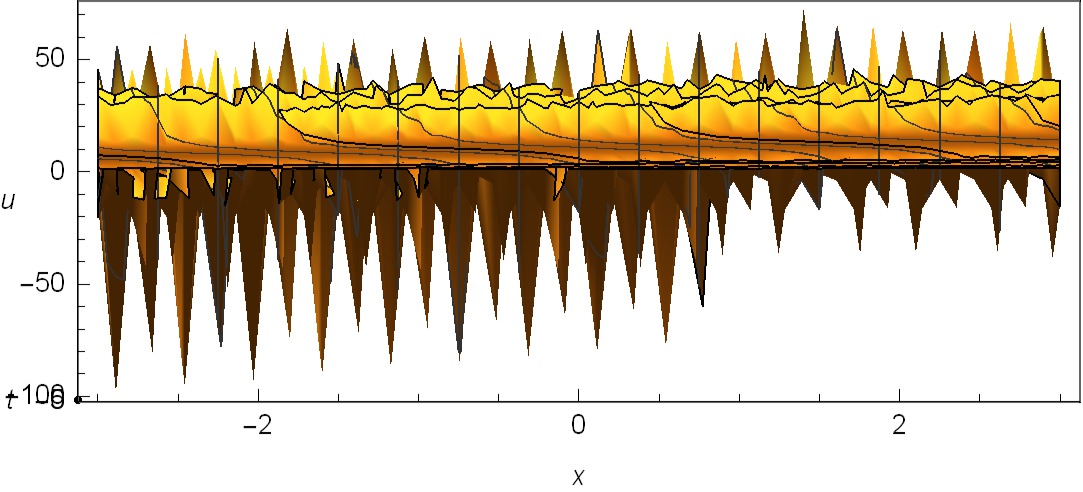}}%
	\hfill
	\subcaptionbox{}{\includegraphics[width=0.50\textwidth]{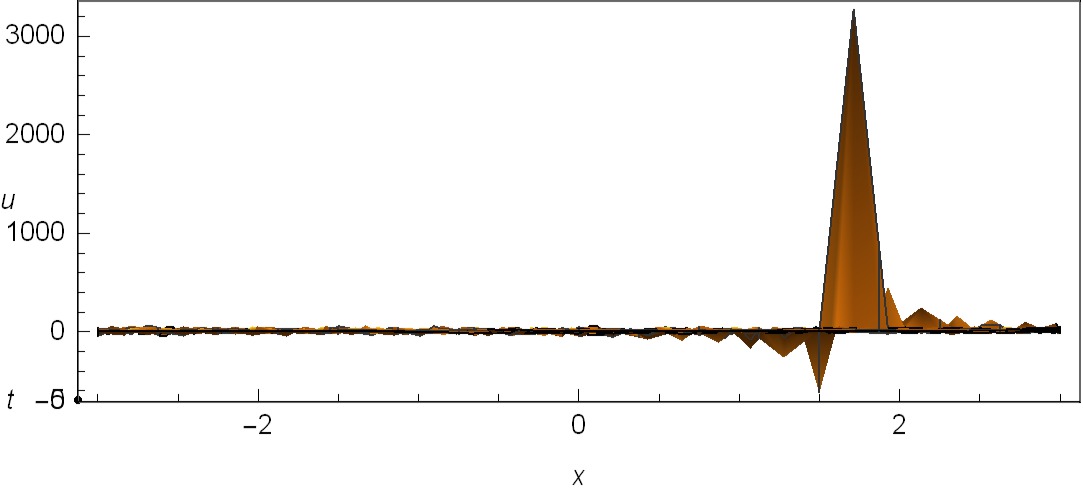}}%
	\caption{{Evolution profiles of 1-Multisoliton solution for Eq. \eqref{solu1a}.}}
	\label{f1c}
\end{figure}

\begin{figure}
	\centering
\subcaptionbox{Progressive wave shown by solution Eq. \eqref{1bi}}{\includegraphics[width=0.50\textwidth]{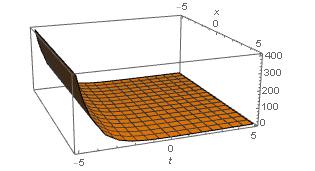}}%
\subcaptionbox{Singularity shown by solution Eq. \eqref{1bii}}{\includegraphics[width=0.50\textwidth]{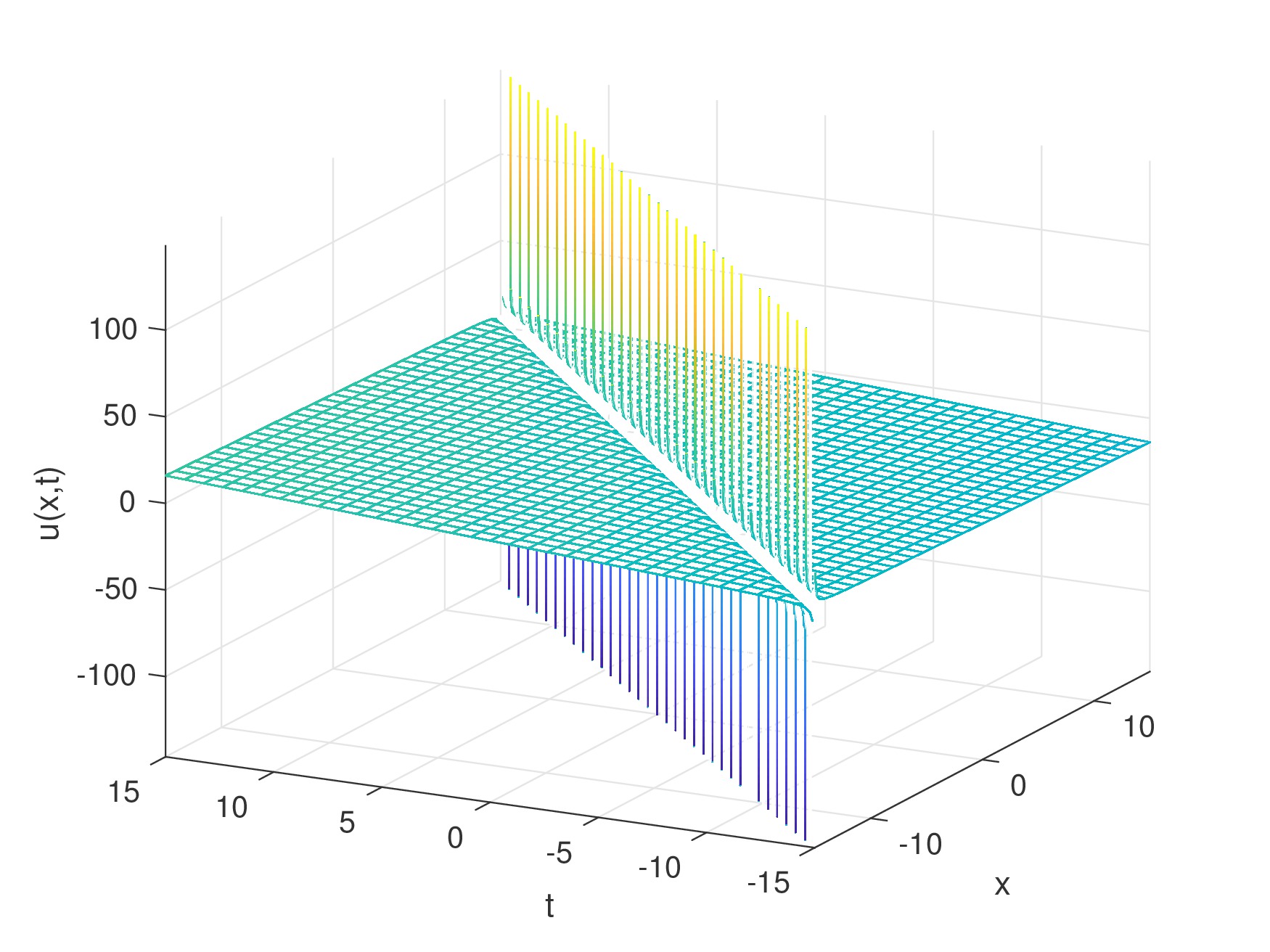}}%
	\hfill
\subcaptionbox{Singularity on $1+x-t=0$ in Eq. \eqref{solu2b}}{\includegraphics[width=0.50\textwidth]{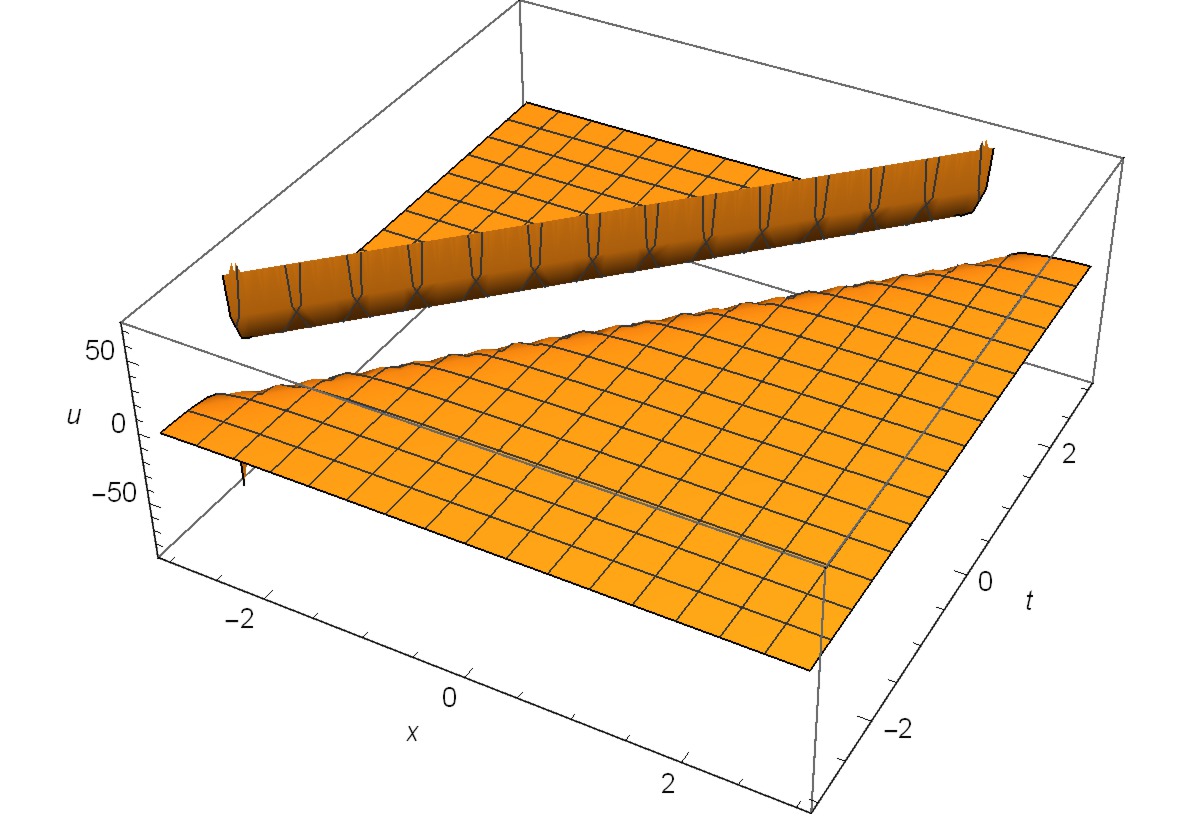}}%
	\hfill
\subcaptionbox{Singularity in the solution Eq. \eqref{solu2b2}}{\includegraphics[width=0.50\textwidth]{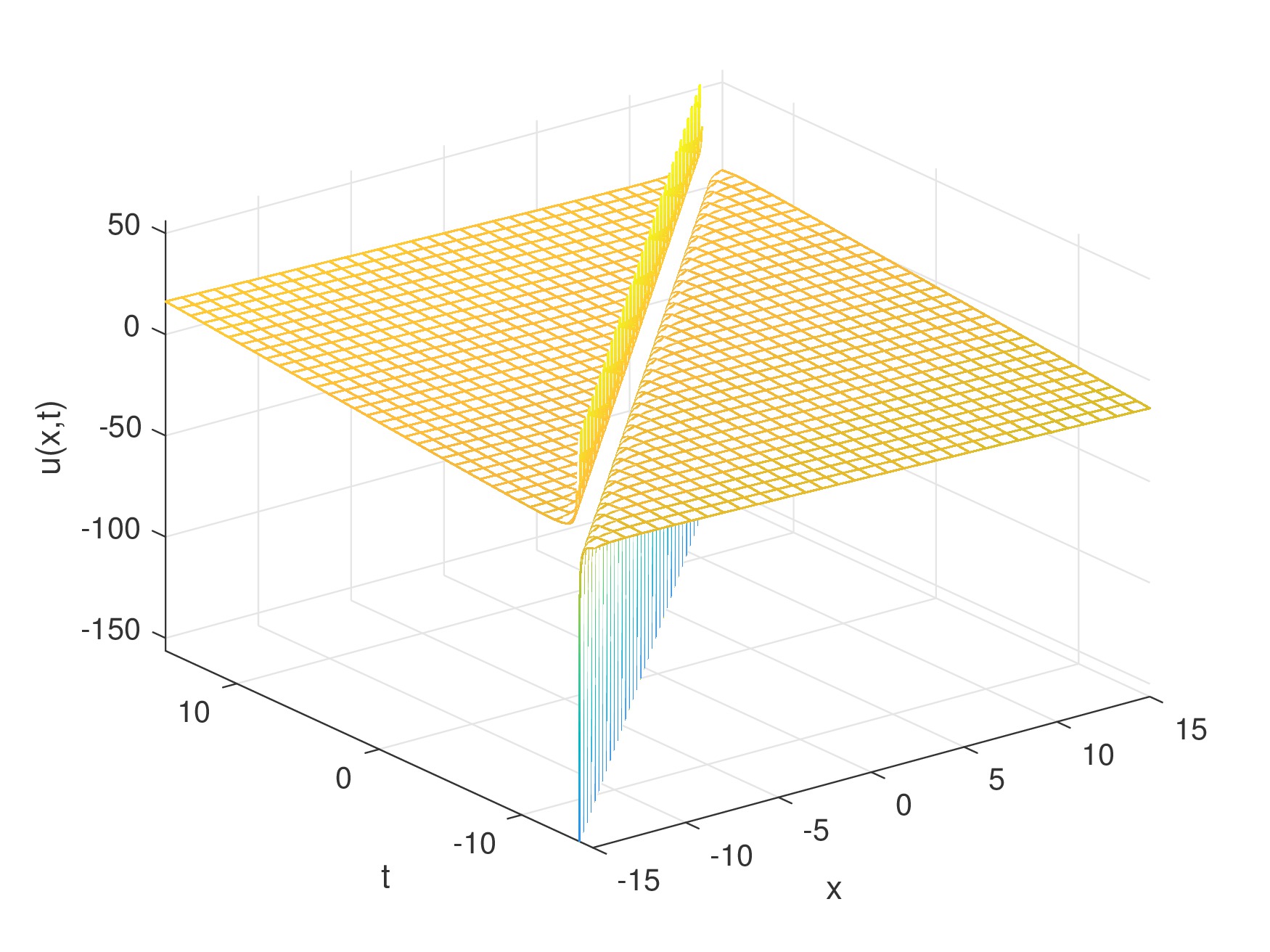}}%
	\hfill
	\caption{Evolution profiles of various invariant solutions.}
	\label{firstfig}
\end{figure}

\begin{figure}[hb]
	\centerline{
		\includegraphics[width=5.5cm]{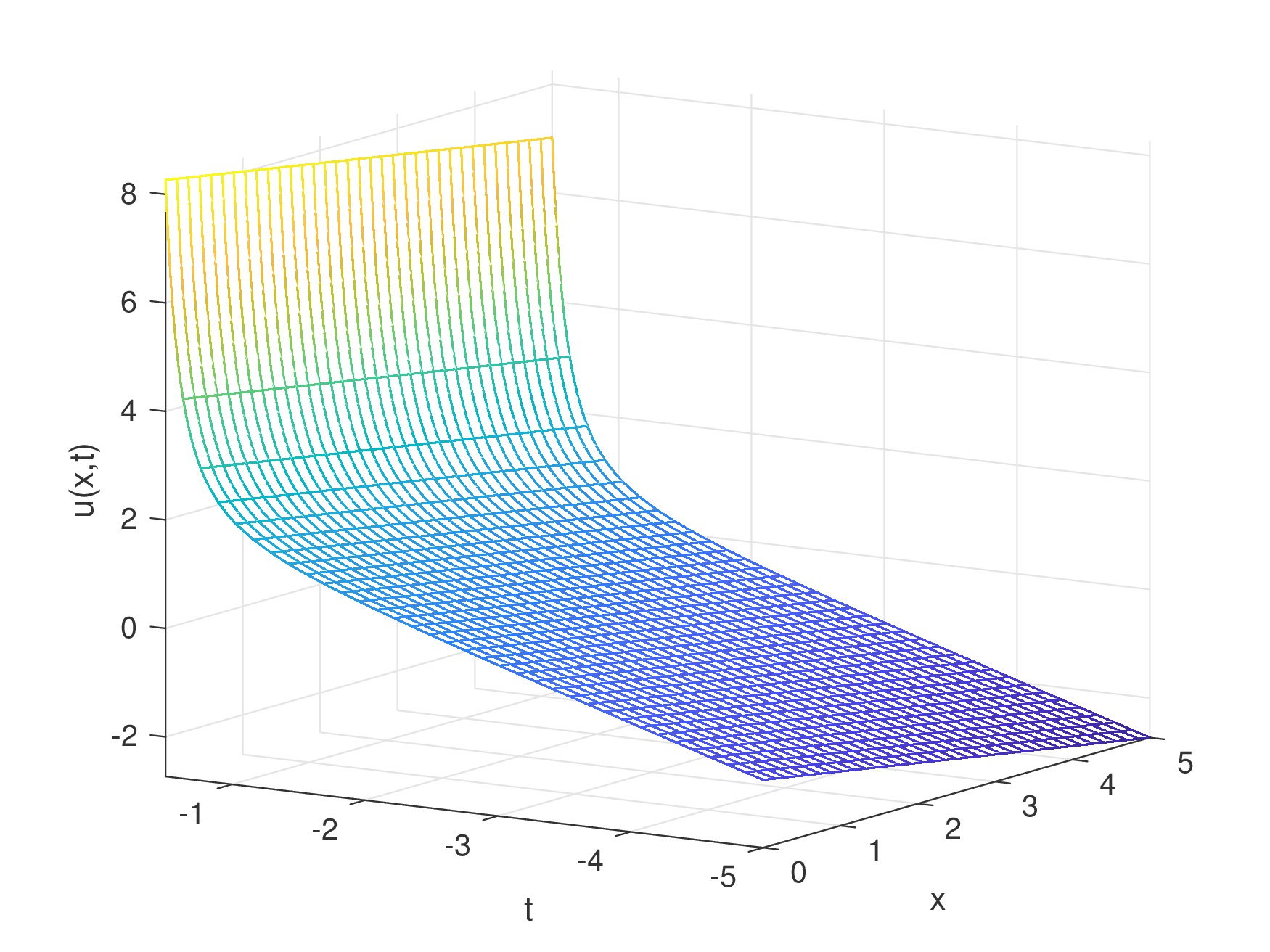}
		\includegraphics[width=5.5cm]{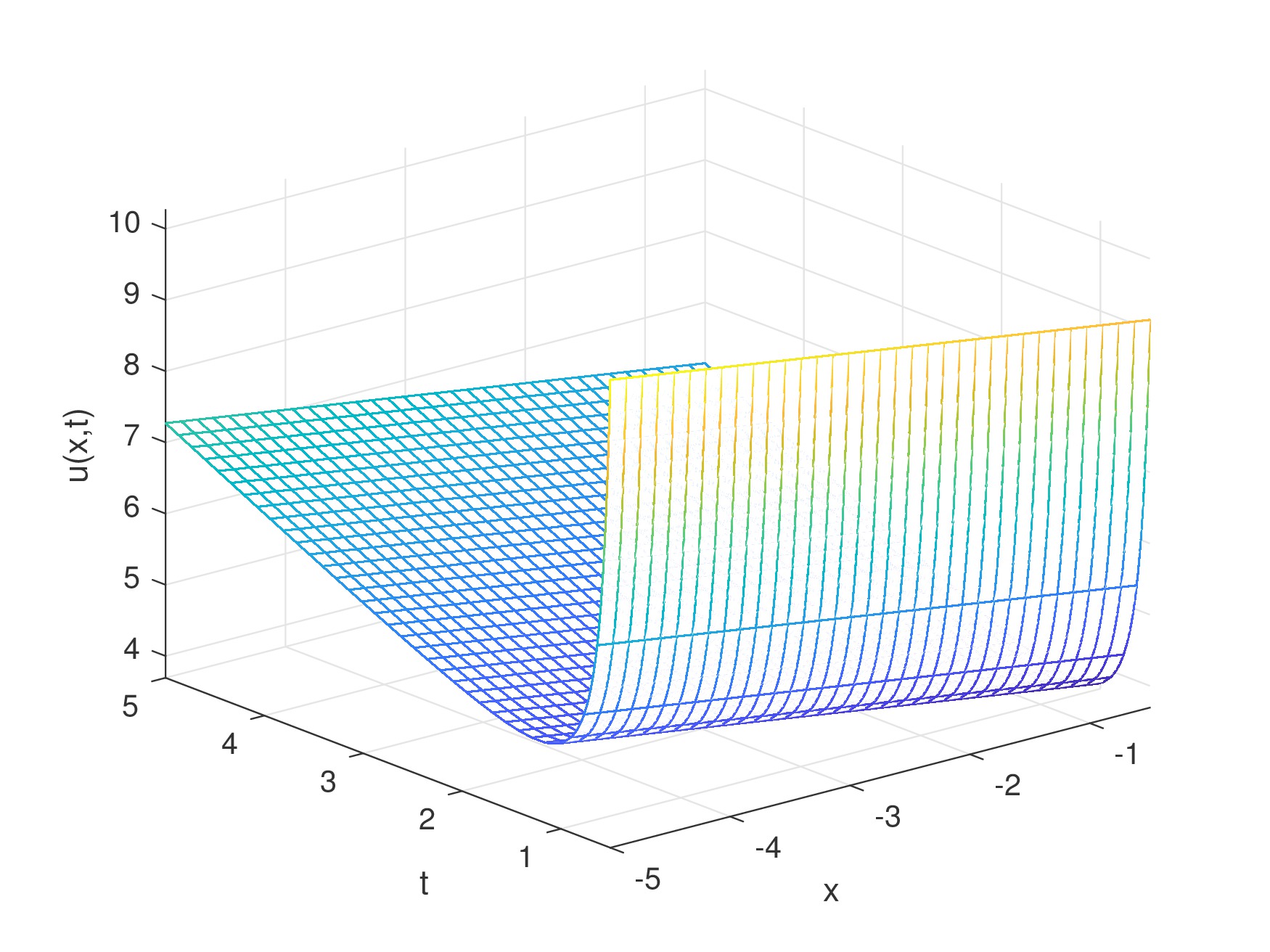}
	}
	\vspace*{8pt}
	\caption{Asymptotic structures at $t=0$ for Eq. \eqref{u3awithc}.}
	\label{fig4}
\end{figure}

The organization of the paper is as follows. In Sec. 2, we discuss the  methodology of Lie symmetry analysis of the general case. In Sec. 3, we obtain infinitesimal generators and the Lie point symmetries of the  Eq. \eqref{eq1}.  
In Sec. 4, symmetry reductions and exact group invariant solutions for the combined KdV-nKdV eqation were obtained. In Sec. 5, we discussed all the invariant solutions graphically by Figures \ref{f1a}, \ref{f1c}, \ref{firstfig} and \ref{fig4}. Finally, concluding remarks are summarized in Section 6.
\section{Method of Lie symmetries}
Let us consider a system of partial differential equations as follows:
\begin{align}\label{sys1}
	\Lambda_\nu (x,u^{(n)})=0,\,\, \nu=1,2,...,l,
\end{align}
where $u=(u^1,u^2,...,u^q), x= (x^1,x^2,...x^p), u^{(n)}$ denotes all the derivatives of $u$ of all orders from $0$ to $n$.
The one-parameter Lie group of infinitesimal transformations for Eq. \eqref{sys1} is given by
\begin{align}
	\tilde{x}^i &=x^i+\epsilon\, \xi^i (x,u) +O(\epsilon^2); i=1,2,...p, \nonumber \\
	\tilde{u}^j &=u^j+\epsilon\, \phi^j (x,u) +O(\epsilon^2); j=1,2,...q, \nonumber
\end{align}
where $\epsilon$ is the group parameter, and  the Lie algebra of  Eq. \eqref{eq1} is spanned by vector field of the form
\begin{align}
	V=\sum_{i=1}^{p} \xi^i (x,u) \frac{\partial}{\partial x^i}+\sum_{j=1}^{q} \eta^j (x,u) \frac{\partial}{\partial u^j}
\end{align}
A symmetry of a partial differential equation is a transformation which keeps the solution invariant in the transformed space. The system of nonlinear PDEs leads to the following invariance condition under the infinitesimal transformations
\begin{align*}
	Pr^{(n)} V [\Lambda_{\nu}(x,u^{(n)})]=0, \nu=1,2,...l \,\,\text{along with } \,\, \Lambda(x,u^{(n)})=0
\end{align*}
In the above condition, $Pr^{(n)}$ is termed as $n^{th}$-order
prolongation \cite{BookOlver} of the infinitesimal generator $V$ which is given by 
\begin{align}
	Pr^{(n)}V=V
	+\sum_{j=1}^{q}\sum_{J}^{} \eta_j^J (x,u^{(n)}) \frac{\partial}{\partial u_J^j}
\end{align}
the second summation being over all (unordered) multi-indices $J=(i_1,...,i_k)$,  $1\le i_k\le p,\,\, 1 \le k \le n$. The coefficient functions $\eta_j^J$ of $Pr^{(n)}V$ are given by the following expression

\begin{align}
	\eta_j^J (x,u^{(n)})=D_J \bigg(\eta_j-\sum_{i=1}^{p} \xi^i u_i^j\bigg)+\sum_{i=1}^{p}\xi^i u_{J,i}^j
\end{align}
where $u_i^j=\frac{\partial u^j}{\partial x^i}$, $u_{J,i}^j=\frac{\partial u_J^j}{\partial x^i}$ and $D_J$ denotes total derivative.

\section{Lie symmetry analysis for the combined KdV-nKdV equation}
In this section, authors explained briefly all the steps of the STM method to keep this work self-confined. The Lie symmetries
for the Eq. \eqref{eq1} have generated and then its similarity solutions are found. Therefore, one can consider the following
one-parameter ($\epsilon$) Lie group of infinitesimal transformations
\begin{align}
\tilde{x} &=x+\epsilon \, \xi (x,t,u) +O(\epsilon^2), \nonumber \\
\tilde{t} &=t+\epsilon \, \tau (x,t,u) +O(\epsilon^2), \nonumber \\
\tilde{u} &=u+\epsilon \, \eta (x,t,u) +O(\epsilon^2),
\end{align}
where $\xi, \tau$ and $\eta$ are infinitesimals for the  variables $x, t$ and $u$ respectively, and $u(x,t)$ is the solution of Eq. \eqref{eq1}.
Therefore, the associated vector field is 
\begin{equation}\label{vectorfield}
{\bf V}=\xi(x,t,u) \frac{\partial}{\partial x}+\tau(x,t,u) \frac{\partial}{\partial t}+\eta(x,t,u) \frac{\partial}{\partial u}.
\end{equation}
Lie symmetry of Eq. \eqref{eq1} will be generated by Eq. \eqref{vectorfield}. Use fourth prolongation $Pr^{(4)} V$ gives rise to the symmetry condition for Eq. \eqref{eq1} as follows: 
\begin{align}\label{deteq}
\eta^{xt}+ 6 \eta^x u_{xx} + 6 \eta^{xx} u_x  +&\eta^{xxxx}+\eta^{xxxt}+4 \eta^x u_{xt}\nonumber \\+&4 \eta^{xt} u_x +2 \eta^t u_{xx} +2 \eta^{xx} u_t=0,
\end{align}
where $\eta^x, \eta^t, \eta^{xt}, \eta^{xx}, \eta^{xxxx}$, and $\eta^{xxxt}$ are the coefficient of $Pr^{(4)}V$,  values are given in many references \cite{BookOlver,STM}. Incorporating all the expressions into Eq. \eqref{deteq}, and then equating the various differential coefficients of $u$ to zero, we derive following system of Eight determining equation
\begin{align}
\xi_u=\xi_{xx}=0, \,\,\,
\xi_x+\xi_t=\tau_t, \,\,\,
\tau_x=\tau_u=0,\nonumber\\
\xi_x+2 \eta_x=\xi_x+\eta_u=0,\nonumber \\
\xi_x+\frac{1}{2} \tau_t=\eta_t.
\end{align}
Solving the above system of equations, we obtain following
infinitesimals for \eqref{eq1} using software {\it Maple},
\begin{eqnarray}{\label{generatorsG}}
\xi  = (x-t)a_1+a_2 +f(t), \,\,
\tau = f(t),\,\,
\eta = (t-\frac{x}{2}-u)a_1+a_3 +\frac{1}{2} f(t),
\end{eqnarray}
where $a_1, a_2$ and $a_3$ are arbitrary constants whereas $f(t)$ is an arbitrary function.

The symmetries under which Eq. \eqref{eq1} is invariant can be spanned by the following four infinitesimal generators if we assume $f(t) = c$, a constant
Then all of the infinitesimal generators of Eq. \eqref{eq1} can be expressed as
\begin{align*}
V=a_1 V_1+a_2 V_2+a_3 V_3+c V_4.
\end{align*}
where 
\begin{eqnarray}\label{Fourvector}
V_1&=&2 (x-t) \frac{\partial}{\partial x} + (t-\frac{x}{2} -u) \frac{\partial}{\partial u}\label{eqv1}, \nonumber \\
V_2&=&  \frac{\partial}{\partial x}\label{eqv2}, \nonumber \\
V_3&=&\frac{\partial}{\partial u} \label{eqv3}, \nonumber \\
V_4&=& \frac{\partial}{\partial x}+\frac{\partial}{\partial t}+\frac{\partial}{\partial u}.
\end{eqnarray}

\begin{table}[ht]
	\centering
	\caption{The commutator table of the vector fields \eqref{Fourvector}}
	{\begin{tabular}{@{}ccccc@{}} 
			\hline
			*   & $V_1$ & $V_2$ & $V_3$ & $V_4$    \\ 
			\hline
			$V_1$ & \hphantom{0} 0 & \hphantom{0}$-V_2+\frac{1}{2} V_3$ & $V_3$&$\frac{1}{2}V_3$ \\
			$V_2$ & $V_2-\frac{1}{2} V_3 $ & 0  & 0&0 \\
			$V_3$  & $-V_3$ & 0      & 0&0  \\ 
			$V_4$  & $-\frac{1}{2}V_3$ & 0      & 0  &0\\ 
			\hline
	\end{tabular}}
	\label{tab1a}
\end{table}

The  vector field yield commutation relations through the Table \ref{tab1a}. The $(i,j)$th  entry in Table \ref{tab1a}  is the Lie bracket 
$[V_i,  V_j] =V_i \cdot V_j-V_j \cdot V_i$. Table 1 is skew - symmetric with zero diagonal elements. 
Table \ref{tab1a} shows that the generators $V_1, V_2, V_3$ and $V_4$ are linearly independent. Thus, to obtain the similarity solutions of
Eq. \eqref{eq1}, the corresponding associated Lagrange system is
\begin{equation} \label{LagrageG}
\frac{dx}{\xi(x,t,u)}=\frac{dt}{\tau(x,t,u)}=\frac{du}{\eta(x,t,u)}.
\end{equation}
\section{Invariant solutions of the combined KdV-nKdV equation}
To proceed further, selection of $f(t)$ and by assigning the  particular values to $a_i$'s $(1 \le i \le 3)$,  provide new physically meaningful solutions of Eq. \eqref{eq1}. In order to obtain symmetry reductions and invariant solutions, one has to solve the associated Lagrange equations given by \eqref{LagrageG}.
Now, let us discuss the following particular cases for various forms of $f(t)$:\\

\noindent {\bf Case 1:} For $f(t)=a\, t^2 +b\, t+ c, a \ne 0, b \ne 0, c\ne 0$, then Eqs. \eqref{generatorsG} and \eqref{LagrageG} gives
\begin{equation} \label{Lagrangef(t)}
\frac{dx}{(x-t)a_1+a_2 +f(t)}=\frac{dt}{f(t)}=\frac{du}{(t-\frac{x}{2}-u)a_1+a_3+\frac{1}{2} f(t)}.
\end{equation}
\noindent The similarity form suggested by Eq. \eqref{Lagrangef(t)} is given by
\begin{equation}\label{uwithX1}
u = \alpha + \frac{1}{4}(3t-x) +e^{-2a_1 \beta \tan^{-1}{(\beta T)}} U(X),
\end{equation}
with similarity variable
\begin{equation}\label{X1}
X=(x-t+A) e^{-2a_1 \beta \tan^{-1}{(\beta T)}},
\end{equation}
where 
\begin{equation}\label{Constants}
\alpha=\frac{a_2+4a_3}{4a_1}, \beta=\frac{1}{\sqrt{4ac-b^2}}, A=\frac{a_2}{a_1},T=2 a t+b.
\end{equation}
Inserting the value of $u$ from Eq. \eqref{uwithX1} into Eq. \eqref{eq1}, we get the following fourth order ordinary differential equation
\begin{equation}\label{ode1}
X \,U_{XXXX} +4 \,U_{XXX}+6\,X\, U_X \,U_{XX} +2\,U \,U_{XX} +8 \,U_{X}^2=0,
\end{equation}
where $X$ is given by Eq. \eqref{X1} and $U_{X}=\frac{dU}{dX}, U_{XX}=\frac{d^2U}{dX^2}$, etc. \\

Any how, we could not find the general solution of Eq. \eqref{ode1} still two particular solutions are found as below
\begin{align}\label{solU1}
U(X) =c_1 \,\,\, \text{and} \,\,\, U(X)=\frac{c_2}{X},
\end{align}
where $c_1$ and $c_2$ are arbitrary constants.
Thus, from Eqs. \eqref{uwithX1} and \eqref{solU1}, we get two invariant solutions  of Eq. \eqref{eq1} given below 
\begin{align}
u(x,t) &= \alpha + \frac{1}{4}(3t-x) +c_1 \,e^{-2a_1 \beta \tan^{-1}{(\beta T)}}, \label{u1awithc} \\
u(x,t)& = \alpha + \frac{1}{4}(3t-x) +\frac{c_2}{A+x-t} \label{u1bwithc},
\end{align}
where $\alpha$, $\beta, \,T$ and $A$ are given by Eq. \eqref{Constants}.\\

\noindent {\bf Case 2:} For $f(t) =c,, c \ne 0; \,\,a_1 =0, a_2 \ne 0, a_3 \ne 0$, Eq. \eqref{LagrageG} are of the form
\begin{equation}\label{Lagrange_a4_a1}
\frac{dx}{c+a_2}=\frac{dt}{c}=\frac{du}{\frac{c}{2}+a_3}.
\end{equation}
The similarity solution to Eq. \eqref{Lagrange_a4_a1} can be written as
\begin{equation}\label{ucase1a}
u=\bigg(\frac{1}{2}+\frac{a_3}{c} \bigg) t+ U(X),
\end{equation}
where $X = x-(1+\frac{a_2}{c}) t$ is a similarity variable.\\
Inserting the value of $u$ from Eq. \eqref{ucase1a} in Eq. \eqref{eq1}, we get the fourth order ordinary differential equation in $U$
\begin{equation}\label{ode2}
a_2 \, U_{XXXX}+(a_2-2 a_3 +6 \, a_2 \,U_X) \,U_{XX}=0.
\end{equation}
Assume $a_2 =2 a_3$. Without loss of generality, we can assume $a_2 \ne 0$, then Eq. \eqref{ode2} reduces to 
\begin{equation}\label{ode3}
U_{XXXX}+6 \, U_X U_{XX}=0.
\end{equation}
The general solution of Eq. \eqref{ode3} in terms of WeiestrassZeta function as 
\begin{equation}\label{Ucase1a}
U(X)= c_3+ (-1)^{\frac{1}{3}} \,2^{\frac{2}{3}} \,WeierstrassZeta[\alpha_1(x,t),\alpha_2],
\end{equation}
where $\alpha_1(x,t)=(-\frac{1}{2})^{\frac{1}{3}} \bigg(x-(1+\frac{a_2}{c}) t+c_4\bigg),\,
\alpha_2=\{2 \,(-1)^{\frac{2}{3}} \, 2^{\frac{1}{3}} c_4, c_5 \}$ and $c_3, c_4$ and $c_5$ are arbitrary constants. 
From Eq. \eqref{Ucase1a} with Eq. \eqref{ucase1a},  we have another invariant solution of Eq. \eqref{eq1} 
\begin{align}\label{solu1a}
u(x,t)=c_3+\bigg(\frac{1}{2}+\frac{a_3}{c} \bigg) t 
+(-1)^{\frac{1}{3}} \,2^{\frac{2}{3}}WeierstrassZeta [\alpha_1(x,t),\alpha_2].
\end{align}


\noindent {\bf Case 2A:} $f(t) = c; a_2 = 0, a_1 \ne 0, a_3 \ne 0 $, Eq. \eqref{Lagrangef(t)} becomes
\begin{equation} \label{Lagrangecase1B}
\frac{dx}{(x-t)a_1 +c}=\frac{dt}{c}=\frac{du}{(t-\frac{x}{2}-u)a_1+a_3+\frac{c}{2}}.
\end{equation}
In this case, we get
\begin{equation}\label{ucase1B}
u=\frac{a_3}{a_1}+\frac{1}{4}(3t-x) +e^{-\frac{a_1}{c} t} U(X),
\end{equation} 
where $X=(x-t)e^{-\frac{a_1}{c} t}$. Substituting the value of $u$ in Eq. \eqref{eq1}, again  we obtain the same fourth order ordinary differential equation  \eqref{ode1}. Some particular solutions are given below
\begin{align}\label{case2U}
U(X)=c_6,\,\,\,\, U(X) =\frac{c_7}{X},
\end{align}
where $c_6$ and $c_7$ are arbitrary constants. Therefore, using Eq. \eqref{case2U} in Eq. \eqref{ucase1B}, we obtain the following two exact solutions for Eq. \eqref{eq1}
\begin{align}
u(x,t)&=\frac{a_3}{a_1}+\frac{1}{4} (3t-x) + c_6 e^{-\frac{a_1}{c} t}, \label{1bi} \\
u(x,t)&=\frac{a_3}{a_1}+\frac{1}{4}(3t-x) +\frac{c_7}{x-t}. \label{1bii}
\end{align}


\noindent {\bf Case 3:} For $f(t)=0; a_1 \ne 0, a_2 \ne 0, a_3 \ne 0$, Eq. \eqref{Lagrangef(t)} modified as
\begin{equation}{\label{LagrangeCase2}}
\frac{dx}{(x-t)a_1+a_2}=\frac{dt}{0}=\frac{du}{(t-\frac{x}{2}-u)a_1+a_3}.
\end{equation}
The group invariant solution is given as
\begin{align}\label{uwithU2}
u&=\frac{a_1 x (x-4 t)-4 a_3 x+4 U(T)}{4 a_1 (t-x)-4 a_2},
\end{align} 
with $T=t$.	Substituting the value of $u$ from Eq. \eqref{uwithU2} into Eq. \eqref{eq1}, we get the following reduced ordinary differential equation
\begin{equation}\label{ode2}
\small{ \left[3 a_1^2 \left(t^2-1\right)-2 a_1 \left(a_2 t-2 a_3 t+2 U(t)\right)-a_2 \left(a_2+4 a_3\right)\right] \left[3 a_1 t-a_2+2 a_3-2 U'(t)\right]=0}.
\end{equation}
Hence, from Eq.\eqref{ode2} we found two values of $U$ given as
\begin{align}
U(T)&=\frac{3 a_1^2 \left(t^2-1\right)-2 \left(a_2-2 a_3\right) a_1 t-a_2 \left(a_2+4 a_3\right)}{4a_1}, \label{U1} \\
U(T)&=\frac{1}{4} (3 a_1 t^2-2 a_2 t+4 a_3 t+4 c_8), \label{U1a}
\end{align}	
where $c_8$ is an arbitrary constant of integration. Here, Eq. \eqref{U1} gives same solution as Eq. \eqref{u1bwithc} with $c_2=\frac{3}{4}$. Also, using Eq. \eqref{U1a} in Eq. \eqref{uwithU2} gives new invariant solution of Eq. \eqref{eq1} as
\begin{align}
u(x,t)=\frac{4 \left(a_3 (t-x)+c_8\right)+a_1 \left(3 t^2-4 t x+x^2\right)-2 a_2 t}{4 a_1 (t-x)-4 a_2}, \label{solu2b}
\end{align}
where $\alpha$ and $A$ are given by Eq. \eqref{Constants}. \\

\noindent	{\bf Case 3A:} For  $f(t) =0; a_2 =0, a_1 \ne 0, a_3\ne0 $, from Eq. \eqref{LagrangeCase2}, we have 
\begin{equation}\label{case1b}
\frac{dx}{(x-t)a_1}=\frac{dt}{0}=\frac{du}{(t-\frac{x}{2}-u)a_1+a_3}.
\end{equation}
Therefore, the similarity transformation method gives
\begin{align}\label{case2b}
u=\frac{ 4 x a_1 - x (x-4 t) a_3-4 U(T)}{4 (a_1 + (x-t) a_3)},
\end{align}
where $T=t$ and $U(T)$ is the  similarity function. Substituting the value of $u$ in Eq. \eqref{eq1} we get reduced ordinary differential equation given as
\begin{align}\label{ode6}
\left[a_3 \left(3 a_3 \left(t^2-1\right)-4 U(t)\right)+2 a_3 a_1 t-5 a_1^2\right] \left[3 a_3 t+a_1-2 U'(t)\right] =0.
\end{align}
Again it gives two values of $U$ as
\begin{align}
U(T)&=\frac{3 a_3^2\,(t^2-1)+2 a_3 a_1 t-5 a_1^2}{4 a_3}, \label{case2bU1} \,\,\,\,\\
U(T)&=\frac{1}{4} (2 a_1 t+3 a_3 t^2+4 c_9), \label{case2bU2}
\end{align}
where $c_9$ is constant of integration. Here, Eq. \eqref{case2bU1} gives same solution as Eq. \eqref{u1bwithc}. Also, using \eqref{case2bU2} with  \eqref{case2b} gives another new invariant solution of \eqref{eq1} as 
\begin{align}
u(x,t)&=\frac{2 a_1 (2x-t)-a_3 \left(3 t^2-4 t x+x^2\right)-4 c_9}{4 \left(a_3 (x-t)+a_1\right)}. \label{solu2b2}
\end{align}

\noindent {\bf Case 4:} For $f(t) =t^3; a_1\ne 0, a_2\ne 0, a_3\ne 0$, from Eq. \eqref{LagrangeCase2} we get similarity form suggested by Eq.\eqref{Lagrangef(t)} is given by
\begin{equation}\label{uwithX4}
u = \alpha + \frac{1}{4}(3t-x) +e^{\frac{a_1}{2 t^2} } U(X),
\end{equation}
with similarity variable $X=(x-t+A) e^{\frac{a_1}{2 t^2}}$ and $A$ is given by \eqref{Constants}.
In this case, we get same ordinary differential equation as Eq. \eqref{ode1},
one particular solution is as follows
\begin{align}\label{3solU2}
U(X)=c_{10},
\end{align}
where $c_{10}$ is arbitrary constant. Thus, from Eqs. \eqref{uwithX4} and \eqref{3solU2} we get another new invariant solution of Eq. \eqref{eq1} given below 
\begin{align}
u(x,t) &= \alpha + \frac{1}{4}(3t-x) +c_{10} \,\mathrm{e}^{\frac{a_1}{2 t^2}}. \label{u3awithc}
\end{align}

\section{Discussion}
The results of the combined KdV-nKdV equation presented in this paper have richer physical structure then earliar outcomes in the literature \cite{wazwazKdVnKdV}. The recorded results are significant in the context of nonlinear dynamics, physical science, mathematical physics etc. The invariant solutions obtained can illustrate various dynamic behaviour due to existence of arbitrary constants. The nonlinear behaviour of the results are analyzed in the following manner:\\
{\it Figure 1}:
Fig 1(a) shows progressive wave soluion in Eq. \eqref{u1awithc} for $a_1=1.1, a_2=0.9, a_3=1.2, a=0.88, b=0.9 \,\, \text{and} \,\,c = 0.9$.  
Fig 1(b) shows presence of singularity in the plane $1+x-t=0$ in Eq. \eqref{u1bwithc} with values $a_1=2.1, a_2=2.1, a_3=3.1\,\, \text{and}\,\, c_2 = 0.9$.\\
{\it Figure 2}: The evolution profiles of 1-Multisoliton solution is given by Eq. \eqref{solu1a} as shown in this Figure. We have recorded the physical nature with variation in parameters.
Fig 2(a) For $a_2 = 1.9866, a_3 = 1.3241, c = 9.8654, c_4 = 1.2341$ and $c_5 = 0.8954$ a plot of WeiestrassZeta function is shown; 
Fig 2(b) For $a_2 = 5, c = 1, a_3 = 1, c_4 = 1$ and $c_5 = 1$ only few solitons are shown; 
Fig 2(c) Front orthographic projection is shown for $a_2 = 1.9866, a_3 = 1.3241, c = 9.8654,  c_4 = 5.2341$ and $c_5 = 4.8954$;
Fig 2(d) Front orthographic projection is shown for $a_2 = 1.767, a_3 = 1.3241, c = 9.8654,  c_4 = 1.2341$ and $c_5 = 0.8954$. \\
{\it Figure 3}:
Fig 3(a) For $a_1=0.8, a_3 = 0.8$ and $c=1$, Eq. \eqref{1bi} exhibits progressive wave; 
Fig 3(b), shows singularity in planes for Eqs. \eqref{1bii} for $a_1=0.8, a_2=2, a_3 = 0.8$ and $c_7=c_8=c_9=1$;
Fig 3(c, d) shows singularity wave profile for Eq. \eqref{solu2b} and Eq. \eqref{solu2b2} which explain the transition of nonlinear behaviour in the form of opposite rotatory folded sheets.\\
{\it Figure 4}:
$u(x,t)$ exhibits singularity near $t=0$ but asymptotic structures is observed near $t=0$ for parameters $a_1=1, a_2=1, a_3=1\,\,\text{and}\,\, c_{10}=1$.

\section{Conclusion}
In this paper, the similarity reductions and invariant solutions for the combined KdV-nKdV are presented. This paper obtained the 1-multisoliton and other invariant solution of the equation. The method that was used to obtain the exact group invariant solutions is the Lie symmetry analysis approach.
 All the solutions are different from earlier work which have been obtained by  Wazwaz \cite{wazwazKdVnKdV}. 
Eventually, the stucture of combined KdV-nKdV equation is an non-trivial one, which can be clearly seen from the graphically results of invariant solutions. The Lie symmetry analysis method extracts the new forms of analytic solutions which are of physical importance such as condensed matter physics and plasma physics.

\section*{Acknowledgment}
The second author sincerely and genuinely thanks Department of Mathematics, SGTB Khalsa College, University of Delhi for financial support.

\end{document}